\documentclass[a4paper,12pt]{article}

\usepackage[left=2.5cm,right=2.5cm,top=2.5cm,bottom=2.5cm]{geometry}
\usepackage{graphicx}
\usepackage{amssymb}
\usepackage{multirow}
\usepackage{amsmath}
\usepackage{psfrag}
\usepackage{setspace}
\usepackage{subcaption}
\usepackage[colorlinks=true,bookmarks=true]{hyperref}
\usepackage{float}
\usepackage{cite}

\hypersetup{citecolor=blue}
\newcommand{\dif}{\mathrm{d}}
\newcommand{\Eqref}[1]{(\ref{#1})}
\newcommand{\half}{\frac{1}{2}}

\newcommand{\brac}[1]{\left(#1 \right)}
\newcommand{\sbrac}[1]{\left[#1\right]}

\begin{document}

\title{{Exact gravitational lensing in conformal gravity and Schwarzschild--de Sitter spacetime}}
\author{{Yen-Kheng Lim\footnote{Email: phylyk@nus.edu.sg}~  and Qing-hai Wang\footnote{Email: qhwang@nus.edu.sg}}\\\textit{Department of Physics, National University of Singapore,}\\\textit{Singapore, 117551}}
\date{\today}
\maketitle

\begin{abstract}
  An exact solution is obtained for the gravitational bending of light in static, spherically symmetric metrics which includes the Schwarzschild--de Sitter spacetime and also the Mannheim--Kazanas metric of conformal Weyl gravity. From the exact solution, we obtain a small bending-angle approximation for a lens system where the source, lens and observer are co-aligned. This expansion improves previous calculations where we systematically avoid parameter ranges that correspond to non-existent null trajectories. The linear coefficient $\gamma$ characteristic to conformal gravity is shown to contribute enhanced deflection compared to the angle predicted by General Relativity for small $\gamma$.
\end{abstract}

\section{Introduction} \label{Introduction}

Since its first discovery in the 1970s, gravitational lensing has become an important observational tool in cosmology and astrophysics. Recently, there has been an ongoing debate as to whether the cosmological constant plays a role in gravitational lensing.

In the Schwarzschild--de Sitter (SdS) spacetime, the terms involving the cosmological constant $\Lambda$ drop out of the equations of motion for null geodesics. Hence, conventional wisdom holds that $\Lambda$ does not play a role in the motion of light in this metric, and therefore does not contribute to lensing. However, Rindler and Ishak \cite{Rindler:2007zz} argued that this may not actually be the case. Taking the cosine of the angle to be the invariant inner product between the photon's spatial 3-velocity and the optic axis, it does indeed depend on the metric functions, and therefore $\Lambda$ as well.

There has yet been no consensus as to whether $\Lambda$ contributes to lensing. While there have since been some supporting arguments in favour of the idea (e.g., \cite{Sereno:2007rm,Bhadra:2010jr}), more recently there have been various arguments against Rindler and Ishak's proposal \cite{Khriplovich:2008ij,Park:2008ih,Simpson:2008jf,Piattella:2015xga,Butcher:2016yrs}. Nonetheless, these counter-arguments do not dispute the validity of the invariant angle calculation in SdS under static coordinates. Instead, the disputes are mostly about how to translate the results in terms of observable quantities \cite{Simpson:2008jf,Arakida:2011ty,Butcher:2016yrs,Lebedev:2016kun}. As the present established value of the cosmological constant is relatively small, it might not be noticeable under other effects related to astrophysical lenses, such as aberration \cite{Simpson:2008jf,Butcher:2016yrs,Lebedev:2016kun} and the ambiguities in defining cosmological distances \cite{Butcher:2016yrs}.

While this matter remains open, there is some certainty that with the small magnitude of $\Lambda$, any discrepancy in observation coming from the cosmological constant is expected to be correspondingly small. However, this is not the case in current observational data. For example, there appears to be significant discrepancies in mass measurements of galaxy clusters from lensing data when compared to measurements obtained from X-ray observations \cite{Allen:1997vc}. These lensing mass estimates were performed using equations derived under standard General Relativity (GR). At galactic and cosmic scales, GR is known to be plagued with the dark-matter and dark-energy problems.

In recent years, conformal Weyl gravity \cite{Weyl1,Weyl2,Bach1921} has attracted considerable interest as a compelling alternative to GR. One of the main appeals of conformal gravity is that the theory provides a potential resolution to the dark-energy and dark-matter problems \cite{Mannheim:2005bfa}. Unlike GR, this theory is possibly renormalisable, thus providing interesting approaches to quantum gravity \cite{Mannheim:2009qi,Mannheim:2011ds}. Therefore it is worth attempting a lensing analysis under this theory.

Using conformal Weyl gravity (CWG), it was argued that the theory is able to produce the effective potential consistent with the observed galactic rotational curves without the need to introduce dark matter \cite{Mannheim:1992vj,Mannheim:1992tr,Mannheim:2010ti,Deliduman:2015vnu}. This feature can already be seen in the spherically symmetric vacuum solution by Mannheim and Kazanas (MK) \cite{Mannheim:1988dj},
\begin{align}
 \dif s^2=-f(r)\dif t^2+f(r)^{-1}\dif r^2+r^2\brac{\dif\theta^2+\sin^2\theta\,\dif\phi^2}, \label{metric0}
\end{align}
with
\begin{align}
f(r)=1-\frac{b(2-3b\gamma)}{r}-3b\gamma+\gamma r-kr^2, \label{metric}
\end{align}
where $b$, $\gamma$, and $k$ are integration constants. This solution bears a strong resemblance to the Schwarzschild--de Sitter solution with an additional linear potential term $\gamma r$ in its lapse function. Thus, physical phenomena under CWG occurring in the regime $\gamma r \ll 1$ would reproduce the traditional observational tests of GR. Therefore, we expect the term $\gamma r$ to be negligible at Solar-System length scales. At galactic-length scales, the $\gamma r$ term should produce the observed flat galactic rotation curves, at least at the qualitative level as far as the (spherically symmetric, vacuum) solution \Eqref{metric0} remains applicable.

A natural question that follows is whether CWG would be able to reproduce or explain other astrophysical or cosmological observations in gravitational lensing.\footnote{Besides gravitational lensing, other observational tests that have been considered include, for instance, radar echo delay \cite{Edery:1997hu} and orbital precession \cite{Sultana:2012qp}.} As mentioned above, observations indicate that lensing appears to be stronger than expected for masses which were determined from X-ray data \cite{Allen:1997vc}. In the GR model, this additional lensing was attributed to the presence of dark matter. The CWG model is then required to produce the observed lensing without invoking the presence of additional mass in order to be consistent with its description of galactic rotation curves. In other words, for the same lens mass, CWG is expected to predict stronger lensing compared to GR.

In this paper, we attempt to provide a unified description of gravitational lensing within the Schwarzschild--de Sitter and the MK spacetimes. The key step in doing so is to obtain an exact solution to the geodesic equations and apply it to the Rindler-Ishak angle procedure. A formula commonly sought after in the literature is the small-bending-angle approximation for a particular lens system in which the source, lens and observer are co-aligned.

Edery and Paranjape \cite{Edery:1997hu} first calculated this deflection angle in the MK metric. Their results were obtained using the usual method of calculating the change in the coordinate angle $\phi$ under geodesic motion. Their results indicate that in order to produce a stronger deflection than the GR prediction, the sign of $\gamma$ has to be the opposite of the value obtained by fitting of the galactic rotation curve. A possible resolution of this discrepancy can be found by making a suitable gauge choice for the metric before calculating the deflection angle \cite{Edery:2001at}. Using the Rindler-Ishak method mentioned above, the lensing angle was recalculated in \cite{Sultana:2010zz}, which again gives the opposite sign of $\gamma$ as expected from the galactic rotation curves. However, Cattani et al. \cite{Cattani:2013dla}, also using the Rindler-Ishak method,  obtained another lensing formula which has the expected sign of $\gamma$, thus negating the need to choose an appropriate gauge 
as described by \cite{Edery:2001at}. The seemingly contradictory results of \cite{Sultana:2010zz} and \cite{Cattani:2013dla} deserves further scrutiny, and is one of the main points to be addressed in this paper.


We will argue that these seemingly contradictory formulas were obtained by possibly erroneous calculations. We note here that the deflection angle formulas of \cite{Sultana:2010zz} and \cite{Cattani:2013dla} were obtained by performing small-mass and $\gamma$ approximations, where the main difference between their two results stem from the different points in the calculations where the higher powers of mass and $\gamma$ are discarded.\footnote{We also note that Edery and Paranjape's results \cite{Edery:1997hu} were also obtained by small-mass and $\gamma$ approximations.} We show that the small parameter expansions must be done with care so as to avoid expanding into a parameter range which corresponds to non-existent solutions for a null trajectory that connects a co-aligned source and observer. Upon obtaining the correct expansions, we explicitly check its agreement with the exact solution.

\par
Other authors have considered lensing by the MK metric without the use of approximations. Villanueva and Olivares \cite{Villanueva:2013gga} solved the geodesic equations exactly to provide the coordinate deflection angle. In their analysis of MK geodesics, Hoseini et al. \cite{Hoseini:2016tvu} provided the deflection angle under the Rindler-Ishak formalism. However, the focus of their work was on the geodesic structure of the MK metric, and it has yet to draw any conclusions as to the sign of $\gamma$ pertaining to lensing observations.

The rest of this paper is organised as follows. In Sec.~\ref{Geodesics} we derive the geodesic equations describing the trajectory of light in the MK metric. We provide a brief review and an alternate derivation for the deflection of light in Sec.~\ref{BendingAngles}, which includes the Rindler-Ishak method. We consider the effect of the cosmological constant and $\gamma$ separately in Sec.~\ref{effect} to see its influence on the bending angle. In Sec.~\ref{ApproximateSolutions} we derive simpler expressions for the bending angle under small-parameter approximations. The paper concludes with a summary and discussion in Sec.~\ref{Conclusion}.

\section{Geodesic equations} \label{Geodesics}

\subsection{Metric and equations of motion}

Throughout this paper, we will take our spacetime to be a static, spherically symmetric metric of the form \Eqref{metric0}.
The (SdS) spacetime is a solution to the Einstein equation with a positive cosmological constant $\Lambda$ with
\begin{align}
 f(r)=1-\frac{2m}{r}-\frac{\Lambda r^2}{3}.
\end{align}
On the other hand, as mentioned in Sec.~\ref{Introduction}, the MK metric of the form \Eqref{metric0} is also a vacuum solution in CWG where $f(r)$ is given by \Eqref{metric}. In the following we will find it convenient to introduce a reparametrisation $b=\frac{1-\sqrt{1-6\gamma m}}{3\gamma}$, so that the MK solution is parametrised in terms of `mass' $m$, where \Eqref{metric} is given by
\begin{align}
 f(r)=\sqrt{1-6m\gamma}-\frac{2m}{r}+\gamma r-kr^2. \label{MK_structure_function}
\end{align}
Clearly, the SdS solution can be recovered as a special case of the MK solution by setting $\gamma=0$ and $k=\frac{\Lambda}{3}$. Thus, in the following it suffices to use \Eqref{MK_structure_function} in our geodesic equations without loss of generality.

The motion of a time-like or null particle is described by a trajectory $x^\mu(\tau)$, where $\tau$ is an appropriate affine parametrisation. The geodesic equations are determined by the Lagrangian $L= \half g_{\mu\nu}\dot{x}^\mu\dot{x}^\nu$, where over-dots denote derivatives with respect to $\tau$. The equations of motion may be derived using the Euler-Lagrange equation $\frac{\dif}{\dif\tau}\frac{\partial L}{\partial\dot{x}^\mu}=\frac{\partial L}{\partial x^\mu}$.
\par
Since $\partial/\partial t$ and $\partial/\partial\phi$ are Killing vectors of the spacetime, we have the first integrals of motion
\begin{align}
 \dot{t}=\frac{E}{f},\quad\dot{\phi}=\frac{\Phi}{r^2\sin^2\theta}, \label{constants}
\end{align}
where $E$ and $\Phi$ are constants of motion, which we may interpret as the energy and angular momentum of the particle, respectively.
\par
The equations of motion for the remaining two coordinates are
\begin{align}
 \ddot{r}&=\frac{f'}{2f^2}\dot{r}^2+rf\dot{\theta}^2-\frac{f'E^2}{2f}+\frac{f\Phi^2}{r^2\sin^2\theta},\label{rddot}\\
 \ddot{\theta}&=-\frac{2\dot{r}\dot{\theta}}{r}+\frac{\cos\theta\,\Phi^2}{r^4\sin^3\theta}, \label{thetaddot}
\end{align}
where primes appearing in $f'$ denote derivatives with respect to $r$. The invariance of the inner product $g_{\mu\nu}\dot{x}^\mu\dot{x}^\nu\equiv\epsilon$ provides a constraint equation
\begin{align}
 \dot{r}^2+r^2f\dot{\theta}^2=E^2-V^2. \label{FirstIntegral}
\end{align}
where $V^2$ is the effective potential given by
\begin{align}
 V^2=\brac{\frac{\Phi^2}{r^2\sin^2\theta}-\epsilon}f.
\end{align}
By appropriately rescaling the parameter $\tau$, the magnitude of $\epsilon$ may be normalised to unity if it is non-zero. Hence, we have for time-like geodesics $\epsilon=-1$ and for null geodesics $\epsilon=0$.
\par
Because of the spherical symmetry of the spacetime, all geodesics can be shown to be confined on a two-dimensional plane. We fix our coordinate system such the plane is at $\theta=\pi/2$ and a constant throughout the motion, so that Eq.~\Eqref{thetaddot} becomes trivial. Furthermore we are interested primarily in null geodesics, where $\epsilon=0$. In this case, the energy and angular momentum always appears in the combination $\frac{E^2}{\Phi^2}$. It follows from Eq.~\Eqref{FirstIntegral} that
\begin{align}
 \frac{E^2}{\Phi^2}=\frac{f(r_0)}{r_0^2},
\end{align}
where $r=r_0$ is the `distance of closest approach', which is the radial position when $\dot{r}=0$. With these considerations, the equations of motion reduce to
\begin{align}
 \brac{\frac{\dif r}{\dif\phi}}^2&=r^4\left[\frac{f(r_0)}{r_0^2}-\frac{f(r)}{r^2}\right]. \label{drdphi}
\end{align}
One can prove that any solution to \Eqref{drdphi} is symmetric about the point $r=r_0$.
\par
In the plane $\theta=\pi/2$, we can find circular photon orbits around the MK spacetime by solving $\frac{\dif(V^2)}{\dif r}=0$. One of these roots give a positive radius, which is
\begin{align}
 r_{\mathrm{ph}}=\frac{1-\sqrt{1-6\gamma m}}{\gamma}\sim 3m\left[1+\frac{3}{2}m\gamma+\frac{9}{2}m^2\gamma^2+\mathcal{O}(m^3\gamma^3)\right]. \label{PhotonSphere}
\end{align}
Thus we see that a positive $\gamma$ results in a larger photon sphere when compared to the Schwarzschild case. Similar to the Scharzschild photon sphere, this circular photon orbit is unstable as we can see that
\begin{align}
 \frac{\dif^2\brac{V^2}}{\dif r^2}\Biggr|_{r=r_{\mathrm{ph}}}=-\frac{2\gamma^4}{\brac{1-\sqrt{1-6\gamma m}}^4}<0.
\end{align}
For the range $0<\gamma<\frac{1}{6m}$, the possible radii of the photon sphere range from $3m<r_{\mathrm{ph}}<6m$.

\subsection{Exact solution for light bending}
To calculate the bending angle, we consider photon trajectories with $\epsilon=0$ where the particle reaches $r=r_0$ at the initial angle $\phi=\phi_0$, as shown in Fig.~\ref{fig_DeflectionDiagram}. (We assume throughout that $r_0$ lies outside the horizon of the spacetime.) Using \Eqref{drdphi}, we may describe light deflection in MK, SdS, or Schwarzschild spacetimes under appropriate choices of parameters for $f$ as given by \Eqref{MK_structure_function}. To find an exact solution it is convenient to introduce the substitution $u=1/r$, so that Eq.~\Eqref{drdphi} becomes
\begin{align}
 \brac{\frac{\dif u}{\dif\phi}}^2&=\frac{r_0\sqrt{1-6m\gamma}-2m+\gamma r_0^2}{r_0^3}-\gamma u-\sqrt{1-6m\gamma}\,u^2+2mu^3,\nonumber\\
   &=2m\brac{u_+-u}\brac{u_0-u}\brac{u-u_-}.
\end{align}
In the second line above we have factorised the third-order polynomial where the roots are given by
\begin{align}
 u_0=\frac{1}{r_0},\quad u_\pm=\frac{r_0\sqrt{1-6m\gamma}-2m\pm\sqrt{r_0^2+2m\gamma r_0^2+4r_0m\sqrt{1-6m\gamma}-12m^2}}{4mr_0}.
\end{align}
Therefore, the equations can be solved by performing the integration
\begin{align}
 \int_{\phi_0}^{\phi}\dif\phi'=\pm\int_{u_0}^{u}\frac{\dif u'}{\sqrt{2m\brac{u_+-u'}\brac{u_0-u'}\brac{u'-u_-}}}. \label{phi_soln}
\end{align}
Because the spacetime is invariant under the reflection $\phi\rightarrow-\phi$, we can, without loss of generality, select the lower sign and evaluate the integral exactly, giving\footnote{The integral in the right-hand side of \Eqref{phi_soln} can be found in 3.131-4, pg.~254 of \cite{gradshteyn2014table}.}
\begin{align}
           \phi(u)&=\phi_0+\frac{2}{\sqrt{2m\brac{u_+-u_-}}}\mathrm{F}\brac{\sin^{-1}\sqrt{\frac{(u_+-u_-)(u_0-u)}{(u_0-u_-)(u_+-u)}},\sqrt{\frac{u_0-u_-}{u_+-u_-}}}, \label{Deltaphi}
\end{align}
where $\mathrm{F}(p,q)$ is the incomplete elliptic integral of the first kind. We can express $u$ as a function of $\phi$ by inverting to obtain
\begin{align}
 u(\phi)&=\frac{u_+(u_0-u_-)\mathrm{sn}\brac{\sqrt{\frac{m(u_+-u_-)}{2}}(\phi-\phi_0),\sqrt{\frac{u_0-u_-}{u_+-u_-}}}^2-u_0(u_+-u_-)}{(u_0-u_-)\mathrm{sn}\brac{\sqrt{\frac{m(u_+-u_-)}{2}}(\phi-\phi_0),\sqrt{\frac{u_0-u_-}{u_+-u_-}}}^2-(u_+-u_-)}, \label{Trajectory_uphi}
\end{align}
where $\mathrm{sn}(p,q)$ is the Jacobi elliptic function of the first kind. Equation \Eqref{Trajectory_uphi} can be verified independently against a numerical integration of \Eqref{rddot} and \Eqref{constants}.

\section{Derivation of the bending angle formula} \label{BendingAngles}

\begin{figure}
 \begin{center}
  \includegraphics[scale=0.95]{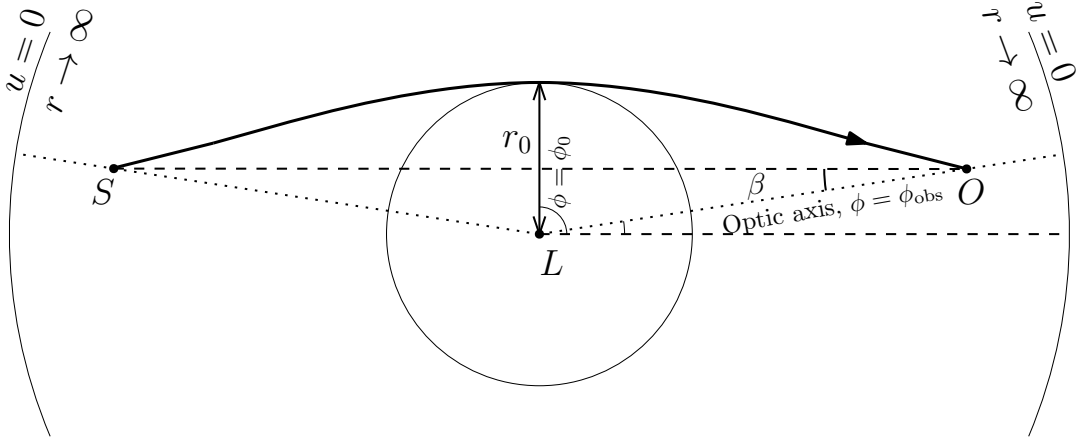}
  \caption{The trajectory of light from source $S$ to observer $O$, passing at a distance of closest approach $r_0$ to the lens $L$.  The asymptotic region $r\rightarrow\infty$ is represented by the outer circular arcs. We assume the trajectory does not cross either the cosmological or event horizons of the spacetime. Here we have drawn the angles $\phi_0$ and $\phi_{\mathrm{obs}}$ to be relative to the horizontal dashed line, implying that this horizontal line is the $\phi=0$ angle. However this is clearly an arbitrary choice and does not affect the analysis.}
  \label{fig_DeflectionDiagram}
 \end{center}
\end{figure}

For a gravitational lens system in SdS and MK spacetime, we consider trajectories depicted in Fig.~\ref{fig_DeflectionDiagram}. As mentioned above, the trajectory is symmetric about the point $r=r_0$, where it begins from a source $S$, passes through the coordinate distance of closest approach $r_0$ to lens $L$, and finally reaches the observer $O$ which we assume to be static with respect to the spatial coordinates of \Eqref{metric}. In a MK spacetime of a given $m$, $k$, and $\gamma$, the possible photon trajectories are parametrised by $r_0$ and are described by the solution \Eqref{Trajectory_uphi}.

Suppose we have an observer located at azimuthal position $\phi_{\mathrm{obs}}$. We define the \emph{optic axis} as the line $\phi=\phi_{\mathrm{obs}}$ that connects the lens to the observer. In Fig.~\ref{fig_DeflectionDiagram}, this is the dotted line $LO$. We then define the observer's position, $r_{\mathrm{obs}}$, as the intersection between the optic axis and the photon trajectory, i.e., $r_{\mathrm{obs}}=1/u(\phi_{\mathrm{obs}})$. For a given $\phi_{\mathrm{obs}}$, we can calculate $r_{\mathrm{obs}}$ accordingly using \Eqref{Trajectory_uphi}.

The difference $\phi_{\mathrm{obs}}-\phi_0$ determines the location of $S$ relative to the optic axis. It is convenient to denote $\beta$ as the angle that parametrises this alignment, defined by
\begin{align}
 \beta=\frac{\pi}{2}-\brac{\phi_{\mathrm{obs}}-\phi_0}. \label{beta_def}
\end{align}
For the special case $\beta=0$, the source, lens, and observer all co-align along the optic axis. By rotating the coordinate system, $\phi_0$ (or $\phi_{\mathrm{obs}}$) can be freely set to any convenient constant. For instance, Ref.~\cite{Rindler:2007zz} sets $\phi_0=\frac{\pi}{2}$, while the analysis in \cite{Edery:1997hu} corresponds to setting $\phi_0=0$. In the following we shall keep $\phi_0$ arbitrary so that our results may accommodate the different conventions.

The observed bending angle depends on the trajectory's (spatial) direction as it arrives at the observer's location. To determine this, let the photon's null 4-velocity be written as $\dot{x}^\mu=(\dot{t},\vec{v})$ where $\vec{v}$ is the space-like component of the 4-velocity, which is further split as
\begin{align}
  \vec{v}&=\vec{v}_\parallel+\vec{v}_\perp;\quad \vec{v}_\parallel=\dot{r}\partial_r,\quad\vec{v}_\perp=\dot{\theta}\partial_\theta+\dot{\phi}\partial_\phi.
\end{align}
When the photon reaches an arbitrary observer at $(t,r,\theta,\phi)$, we may define a local Euclidean orthonormal frame at that location as follows:
\begin{align}
 \vec{e}_{(r)}=\sqrt{f(r)}\partial_r,\quad\vec{e}_{(\theta)}=\frac{1}{r}\partial_\theta,\quad\vec{e}_{(\phi)}=\frac{1}{r\sin\theta}\partial_\phi.
\end{align}
The celestial sphere \cite{Frittelli:1999yf} of the observer is parametrised by angles $\psi$ and $\eta$, where
\begin{align}
 \cos\psi=\vec{e}_{(r)}\cdot\frac{\vec{v}}{|\vec{v}|},\quad \cos\eta=\vec{e}_{(\phi)}\cdot\frac{\vec{v}_\perp}{|\vec{v}_\perp|}.\label{CelestialAngles}
\end{align}
In the above equation, the dot products are understood as the inner product $\vec{a}\cdot\vec{b}=g_{ij}a^ib^j$, and $|\vec{a}|=\sqrt{g_{ij}a^ia^j}$, where the indices $i,\,j$ run along the space-like coordinates. Here we see that $\psi$ is the `polar angle' of the observer's orthonormal frame, or equivalently, the angle between $\vec{v}$ and the optical axis. We note that $\eta$ is the `azimuthal angle' of the observer's orthonormal frame, though this angle is not important for the purposes of the present paper.
\par
Using \Eqref{CelestialAngles} together with \Eqref{FirstIntegral}, we can derive an expression for $\psi$,
\begin{align}
 \frac{r(\phi)\sin\psi}{\sqrt{f(r(\phi))}}=\frac{r_0}{\sqrt{f(r_0)}}.
\end{align}
The angle $\psi$ measured by our stationary observer $O$ located at $\phi=\phi_{\mathrm{obs}}$ is calculated as
\begin{align}
 \sin\psi=\left.\frac{r_0\sqrt{f(r(\phi))}}{r(\phi)\sqrt{f(r_0)}}\right|_{\phi=\phi_{\mathrm{obs}}}=\left.\frac{u(\phi)}{u_0}\sqrt{\frac{f(1/u(\phi))}{f(1/u_0)}}\right|_{\phi=\phi_{\mathrm{obs}}}.\label{InvAngle1}
\end{align}
We can also derive another equivalent formula for $\psi$ by substituting \Eqref{drdphi} into \Eqref{InvAngle1},
\begin{align}
 \sin\psi=\left.\sqrt{\frac{f(1/u(\phi))}{f(1/u(\phi))+\frac{1}{u(\phi)^2}u'(\phi)^2}}\right|_{\phi=\phi_{\mathrm{obs}}},\label{InvAngle2}
\end{align}
where we have denoted $u'(\phi)=\frac{\dif u(\phi)}{\dif\phi}$. This alternate expression, up to trivial applications of trigonometric identities, is precisely the form originally provided by Rindler and Ishak \cite{Rindler:2007zz}, and was used by \cite{Sultana:2010zz} and \cite{Cattani:2013dla} to calculate bending in the MK spacetime in the small $m$ and $\gamma$ regime.
\par
The total bending angle $\hat{\alpha}$ is defined to be equal to $2\psi$. Although Eqs.~\Eqref{InvAngle1} and \Eqref{InvAngle2} are equivalent to each other, the former is more convenient to use because it does not involve any derivatives. Thus for the rest of the paper we will calculate the bending angle using the formula
\begin{align}
 \hat{\alpha}\equiv 2\psi=2\sin^{-1}\left.\sqrt{\frac{u(\phi)^2f(1/u(\phi))}{u_0^2f(1/u_0)}}\right|_{\phi=\phi_{\mathrm{obs}}}, \label{RIangle}
\end{align}
where no approximation has been made, giving the exact bending angle for a spacetime of any $m$, $k$, and $\gamma$.

At this stage, it is important to note that Eq.~\Eqref{RIangle} only holds for $u(\phi_{\mathrm{obs}})>u_{\mathrm{h}}\geq0$, where
$r=1/u_{\mathrm{h}}$ is the location of the cosmological horizon [$f(1/u_{\mathrm{h}})=0$], and the latter inequality is saturated when $k\rightarrow0$ (corresponding to the removal of the cosmological horizon). This condition is equivalent to the statement that a null geodesic passing through $r_0$ is able to intersect the optic axis before crossing beyond the cosmological horizon. Because it is this intersection that defines the location of the observer $u(\phi_{\mathrm{obs}})$, our lensing system is valid with the source and observer being causally connected.


\section{Effect of parameters \texorpdfstring{$k$}{k} and \texorpdfstring{$\gamma$}{gamma} on the bending angle} \label{effect}

With the exact expression \Eqref{RIangle}, we will demonstrate explicitly in this section that the presence of $k$ introduces a diverging effect on a lens system of mass $m$. Furthermore, we will also demonstrate that the conformal gravity parameter $\gamma$ enhances the lensing for small values, while it reduces lensing for larger values. For concreteness, we shall focus on the case $\beta=0$, or equivalently, $\phi_0-\phi_{\mathrm{obs}}=\frac{\pi}{2}$. As described in the previous section, this corresponds to the case where the source, lens, and observer are co-aligned.

\subsection{Lensing in the Schwarzschild--de Sitter metric}

\begin{figure}
 \begin{center}
  \includegraphics{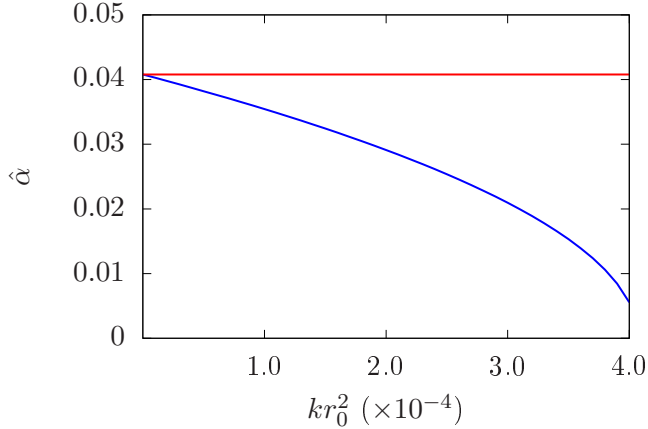}
  \caption{(Colour online.) Plot of the bending angle $\hat{\alpha}$ vs.~$k$ in units of $r_0$, for $m=0.01r_0$, $\gamma=0$, and $\beta=0$. The horizontal red line corresponds to the Schwarzschild deflection value.}
  \label{fig_CompareAlpha_k}
 \end{center}
\end{figure}

We begin with the case $\gamma=0$, corresponding to the Schwarzschild--de Sitter case. For a given $\phi_0$ and $m$, a typical behaviour of $\hat{\alpha}$ is shown in Fig.~\ref{fig_CompareAlpha_k} for varying values of $k$. The parameter $r_0$ can be fixed as the length scale of the system. From the figure, we easily see that any $k\neq0$ gives a deflection less than its pure Schwarzschild ($k=0$) counterpart, showing how a cosmological constant defocuses light passing near the lens.

\begin{figure}
 \begin{center}
  \includegraphics[scale=1]{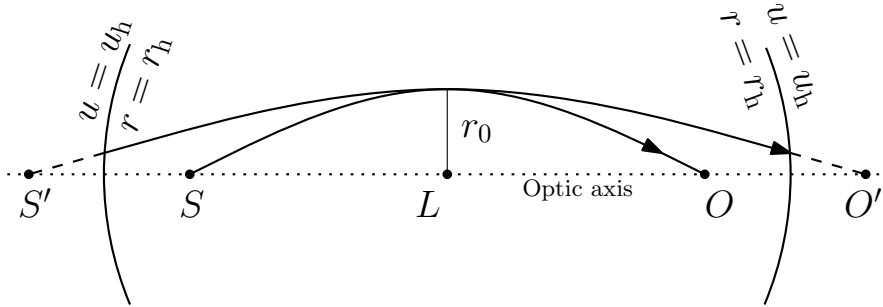}
  \caption{A sketch of trajectories with $m>m_{\mathrm{crit}}$ (the path $SO$) and $m<m_{\mathrm{crit}}$ (the path $S'O'$), for fixed $r_0$. If $m>m_{\mathrm{crit}}$, there exists a trajectory passing through $r_0$ that intersects the optic axis within the cosmological horizon. On the other hand, for $m<m_{\mathrm{crit}}$, the trajectory passing through $r_0$ experiences less bending, and does not intersect the optic axis before crossing the cosmological horizon.}
  \label{fig_CriticalDeflection}
 \end{center}
\end{figure}

For fixed $r_0$, if $m$ is decreased relative to $k$, the bending effect continues to diminish. Therefore, trajectories that pass through $r_0$ intersects the optic axis further from the lens and closer to the cosmological horizon. There will be a critical value $m_{\mathrm{crit}}$ where the trajectory intersects the optic axis precisely on the horizon. Further decreasing the mass beyond $m<m_{\mathrm{crit}}$ the trajectory will not be able to intersect the optic axis before crossing the horizon. This is depicted by the curve $S'O'$ in Fig.~\ref{fig_CriticalDeflection}. In such a case, there is no path connecting a source and observer which are co-aligned along the optic axis.

Because $k$ determines the location of the horizon, the critical value $m_{\mathrm{crit}}$ can be regarded as a function of $k$. The explicit relation is hard to obtain. At small $kr_0^2$, an approximate relation between $m_{\mathrm{crit}}$ and $k$ is found to be
\begin{align}
 \frac{m_{\mathrm{crit}}(k)}{r_0} &\sim \half\sqrt{k}r_0-\brac{-\frac{1}{2} + \frac{15}{64}\pi}kr_0^2 + \brac{-\frac{75}{128}\pi+\frac{225}{1024}\pi^2}k^{3/2}r_0^3 + \mathcal{O}(k^2r_0^4).
\label{eqn:Mcritk}
\end{align}

\begin{figure}
 \begin{center}
  \includegraphics{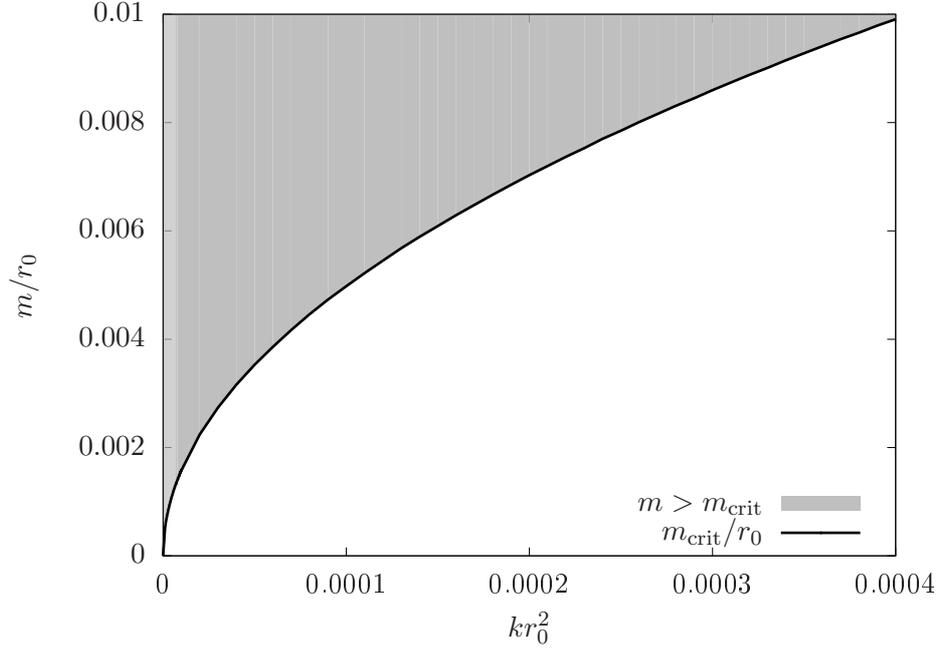}
  \caption{Plot of $m/r_0$ vs.~$kr_0^2$. The shaded region shows the range of parameters where the deflected light is able to reach the observer at $r(\phi_{\mathrm{obs}})<r_{\mathrm{h}}$.}
  \label{fig_k}
 \end{center}
\end{figure}

Figure \ref{fig_k} shows the dependence of $m_{\mathrm{crit}}$ on $k$. The shaded regions are the range of parameters where $m>m_{\mathrm{crit}}$ where trajectories exist for a source, lens and observer are co-aligned along the optic axis. As $k$ increases, $m_{\mathrm{crit}}$ increases monotonically. We can interpret this as the increase of $k$ bringing the horizon closer to the lens, and, hence, a larger mass is required to bend the light towards the optic axis before it crosses the horizon.

\subsection{Lensing in the MK metric with \texorpdfstring{$k=0$}{k=0}}

\begin{figure}
 \begin{center}
  \includegraphics{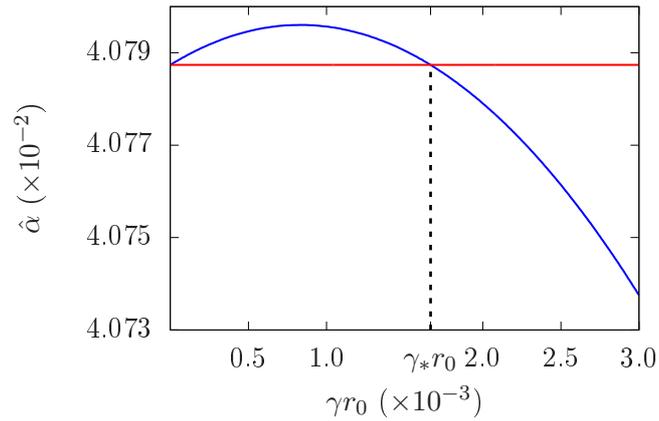}
  \caption{(Colour online.) Plot of $\hat{\alpha}$ vs.~$\gamma r_0$, for $m=0.01r_0$, $k=0$, and $\beta=0$. The horizontal red line corresponds to the Schwarzschild deflection value. In this case, we find that the deflection is greater than the Schwarzschild case for the range $0<\gamma<\gamma_*$, where $\gamma_*\simeq1.6657\times 10^{-3}r_0^{-1}$.}
  \label{fig_CompareAlpha_gam}
 \end{center}
\end{figure}

Turning to the case $k=0$, we now consider the effect of the parameter $\gamma$ on the bending angles. For a given $\phi_0$ and $m$, a typical behaviour of $\hat{\alpha}$ is shown in Fig.~\ref{fig_CompareAlpha_gam} for varying $\gamma  r_0$. As before, the parameter $r_0$ can be fixed as the length scale of the system. For values of $\gamma$ from zero up to a certain $\gamma_*$, the angle $\hat{\alpha}$ is greater than the Schwarzschild value $\hat{\alpha}_{\mathrm{Sch}}$ (the horizontal line in Fig.~\ref{fig_CompareAlpha_gam}), thus giving the result that conformal gravity predicts larger deflection at the range of $0<\gamma<\gamma_*$. The plot shown in Fig.~\ref{fig_CompareAlpha_gam}, shows the results for the choice $m=0.01r_0$, $\beta=0$, and $k=0$. For these parameters, this gives $\gamma_*\simeq1.6657\times 10^{-3}r_0^{-1}$. If $\gamma$ is increased further until a certain value $\gamma_{\mathrm{crit}}$, the bending diminishes until $u(\phi_{\mathrm{obs}})=0$, implying that the trajectory only intersects the optic 
axis at $r\rightarrow\infty$. For the parameters of Fig.~\ref{fig_CompareAlpha_gam}, $\gamma_{\mathrm{crit}}r_0\simeq0.042\,083\,123$.

\begin{figure}
 \begin{center}
  \includegraphics{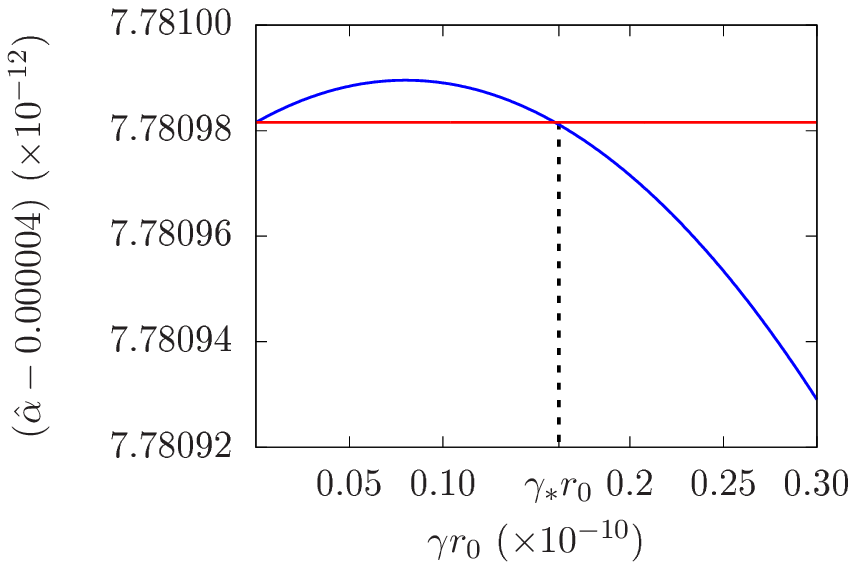}
  \caption{(Colour online.) Plot of $\hat{\alpha}-0.000004$ vs.~$\gamma r_0$, for $m=10^{-6}r_0$, $k=0$, and $\beta=0$. The horizontal red line corresponds to the Schwarzschild deflection value. Here, the value $0.000004$ is subtracted out of $\hat{\alpha}$ to show the numerical variation of $\hat{\alpha}$ more clearly, i.e., the Schwarzschild deflection angle here is $\hat{\alpha}_{\mathrm{Sch}}=0.000\,004\,000\,007\,780\,982$. We find that the deflection is greater than the Schwarzschild case for the range $0<\gamma<\gamma_*$, where $\gamma_*\simeq1.62\times 10^{-10}r_0^{-1}$.}
  \label{fig_CompareAlpha_gamnew}
 \end{center}
\end{figure}

In Fig.~\ref{fig_CompareAlpha_gamnew}, we provide a plot similar to the above, but for the case $m=10^{-6}r_0$. This smaller lens mass corresponds to a more realistic range of parameters corresponding to galaxies and galaxy clusters. The qualitative behaviour of the bending angle is similar to the case $m=0.01r_0$, but the scale is much smaller. In this case, we find that $\gamma_*\simeq1.62\times 10^{-10}r_0^{-1}$.

In the previous paragraphs, we see that the quantities $\gamma_*$ and $\gamma_{\mathrm{crit}}$ changes according to $m$. We could get a better intuitive interpretation by inverting our description such that for a given $\gamma$, there are two quantities $m_*$ and $m_{\mathrm{crit}}$ which we regard as a function of $\gamma$. If $m>m_*$ we have enhanced deflection that results in a bending angle larger than the Schwarzschild angle $\hat{\alpha}_{\mathrm{Sch}}$, in accordance with the expectation of conformal gravity replacing the need for dark matter. However for $m_{\mathrm{crit}}<m<m_*$, we get reduced deflection compared to $\hat{\alpha}_{\mathrm{Sch}}$. At $m=m_{\mathrm{crit}}$, the bending is diminished such that the trajectory passing through $r_0$ could only intersect the optic axis at the infinity (or on the cosmological horizon for the case with $k>0$). For small $\gamma r_0$, the $m_{\mathrm{crit}}$ is found to have the asymptotic behaviour
\begin{align}
 \frac{m_{\mathrm{crit}}(\gamma)}{r_0} \sim \frac{1}{4}\gamma r_0 - \brac{\frac{1}{8}+\frac{15}{256}\pi}\gamma^2r_0^2 + \brac{\frac{1}{16}+ \frac{15}{1024}\pi + \frac{225}{8192}\pi^2}\gamma^3r_0^3 +\mathcal{O}(\gamma^4 r_0^4).
\label{eqn:Mcritg}
\end{align}
If the mass is further decreased beyond $m<m_{\mathrm{crit}}$, the trajectory no longer intersects the optic axis before crossing the horizon. Typical trajectories corresponding to $m>m_{\mathrm{crit}}$ and $m<m_{\mathrm{crit}}$ can be depicted in a sketch similar to Fig.~\ref{fig_CriticalDeflection}, with $u=u_{\mathrm{h}}$ replaced by $u=0$.

\begin{figure}
 \begin{center}
  \includegraphics{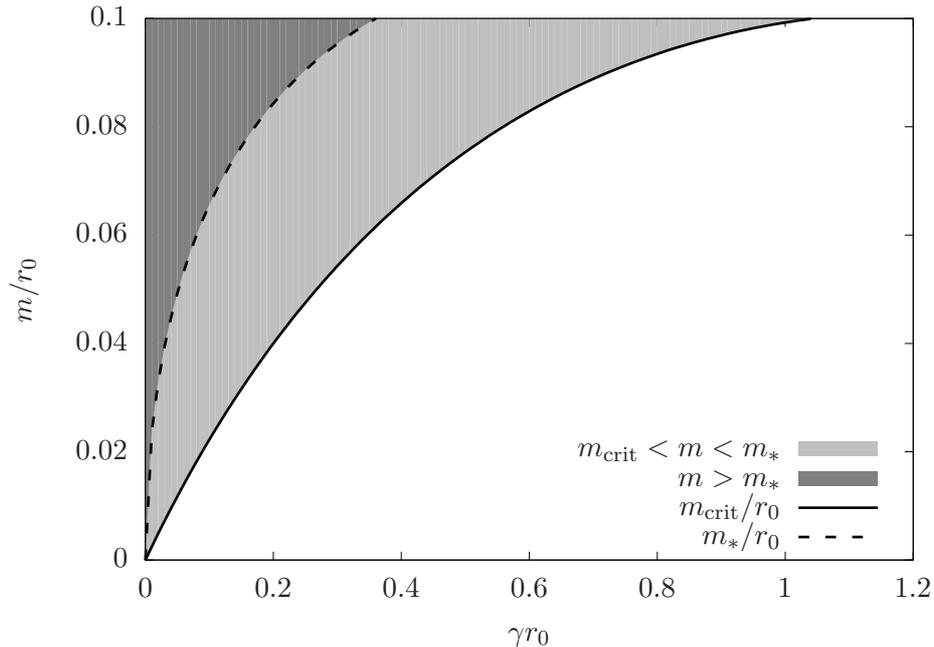}
  \caption{Plot of $m/r_0$ vs.~$\gamma r_0$, where $k=0$. The shaded regions shows the range $m_{\mathrm{crit}}<m<m_*$ which gives reduced deflection, while the darker-shaded regions correspond to $m>m_*$ which gives enhanced deflection.}
  \label{fig_gamstar}
 \end{center}
\end{figure}

\section{Approximate solutions} \label{ApproximateSolutions}

While our exact expression for the bending angle is applicable for a wide range of $m$, $k$, and $\gamma$, its behaviour is buried under various trigonometric and elliptic functions.  In this section we will find  a perturbative expression that allows us to see clearly the relationship between $\hat{\alpha}$ and, say, $\gamma$ without having to resort to numerical exploration.
It would therefore be useful to find a perturbative expansion for $\hat{\alpha}$ for small spacetime parameters.

\subsection{Lensing in the Schwarzschild--de Sitter spacetime}

We first consider small-angle approximations for bending in the Schwarzschild--de Sitter spacetime. We set $\gamma=0$ and attempt to expand \Eqref{RIangle} in small $m/r_0$ and $kr_0^2$. For a given $kr_0^2$, the critical value $m_{\mathrm{crit}}/r_0$ constitutes a lower bound of the lens mass such that a lensing event can take place for a co-aligned source-lens-observer system. This gives us an indication that if we were to find a perturbative expression of $\hat{\alpha}$ for a given co-aligned system, expanding about $m=0$ is ill defined. This is because perturbing about $m=0$, or more specifically any $m<m_{\mathrm{crit}}$, implies an expansion about a non-existent trajectory!

In Ref.~\cite{Rindler:2007zz}, the photon trajectories were expanded about a `straight line' $r\sin\phi=\mathrm{constant}$, but for an arbitrary $\beta$. Thus, the Rindler-Ishak bending angle remains well defined even with the zeroth-order straight-line solution. However, to obtain a perturbative expression showing a leading-order contribution of lensing due to the cosmological constant, a small $m$ and $k$ expansion was made, and $\beta$ was set to zero. Hence, there was a tacit assumption that the observer is located at the `region of transition between Schwarzschild and de Sitter geometry' \cite{Rindler:2007zz}. Thus, any contribution coming from $\Lambda$ (or $k$) is due to the small influence creeping in from the de Sitter side.

In using their approximate expression for $\hat{\alpha}$, this assumption has to be enforced by hand. For a given trajectory specified by $r_0$ deflected by a mass $m$, one has to ensure that the corresponding choice of $k$ does not result in $r(\phi_{\mathrm{obs}})$ being located beyond the horizon. Furthermore, if $m$ and $k$ are treated as independent variables and expanded separately, the result might violate the above assumption if $k$ is not chosen appropriately.

With our exact trajectory \Eqref{Trajectory_uphi} and bending angle \Eqref{RIangle}, we can build in a consistent machinery to ensure the existence of  a trajectory that connects a source co-aligned with the observer. The parameter space that allows such trajectories is represented by the shaded region in Fig.~\ref{fig_k}.  Performing a small-$m$ expansion means that we are expanding about a small neighbourhood around $m=0$. (This would be the small region close to the origin of Fig.~\ref{fig_k}.) That neighbourhood consists of two regions separated by a curve $m=m_{\mathrm{crit}}$. The region $m>m_{\mathrm{crit}}$ is the shaded region where a null trajectory intersects the optic axis before crossing the horizon. Conversely, for the other region $m<m_{\mathrm{crit}}$, there is no trajectory that intersects the optic axis before crossing the horizon, hence an observer will not be able to see a source that lies on the optic axis.

To remain within that region when performing a small $m$ expansion, we note that $k$ has to diminish at a rate fast enough so that $m$ does not overtake $m_{\mathrm{crit}}(k)$. From the asymptotic behaviour of $m_\mathrm{crit} (k)$ in \Eqref{eqn:Mcritk}, we learn that $k$ must diminish at the rate of at least $k\propto m^2$. In light of this, we reparameterise $k$ by setting
\begin{align}
 kr_0^2\equiv \kappa\frac{m^2}{r_0^2}. \label{ksub}
\end{align}
Physically, we do expect $k$ to be independent of $m$, and this is reflected by the independence of $\kappa$. However, expressing $k$ using \Eqref{ksub} and considering $\kappa$ to be of order $\mathcal{O}(1)$ or less ensures that the parameters lie within the regime $m>m_{\mathrm{crit}}$, and we remain within the shaded region of Fig.~\ref{fig_k}.

With this parametrisation, we substitute \Eqref{ksub} into \Eqref{RIangle} and expand in powers of $m/r_0$. The result is, up to third order in $m/r_0$,
\begin{align}
 \hat{\alpha}&\sim 2\sqrt{4-\kappa}\frac{m}{r_0}+\frac{1}{\sqrt{4-\kappa}}\brac{-8+\frac{15}{2}\pi - 2\kappa}\frac{m^2}{r_0^2}\nonumber\\
             &\quad + \frac{1}{\brac{4-\kappa}^{3/2} }\sbrac{\frac{784}{3} - 60\pi  - \brac{88-30\pi+\frac{225}{64}\pi^2} \kappa - \kappa^2 + \frac{2}{3}\kappa^3}\frac{m^3}{r_0^3} \nonumber\\
             &\quad+\mathcal{O}(m^4/r_0^4), \label{alphak1}
\end{align}
Clearly the above expansion only holds for $\kappa<4$, which is consistent with the leading behaviour in \Eqref{eqn:Mcritk} as well as our requirement that $\kappa\lesssim\mathcal{O}(1)$.

\begin{figure}
 \begin{center}
  \begin{subfigure}[b]{0.49\textwidth}
   \includegraphics[scale=0.8]{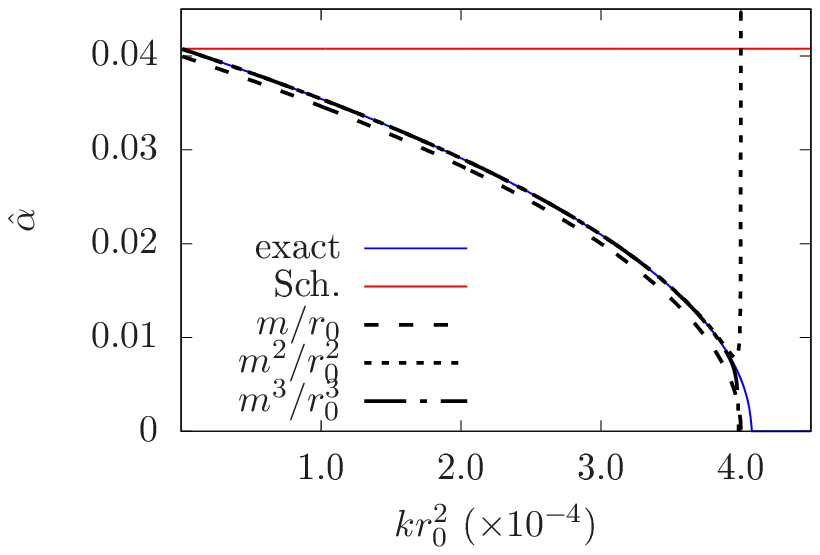}
   \caption{$\hat{\alpha}$ from Eq.~\Eqref{alphak1}.}
   \label{fig_anglesk_compare1}
  \end{subfigure}
  \begin{subfigure}[b]{0.49\textwidth}
   \includegraphics[scale=0.8]{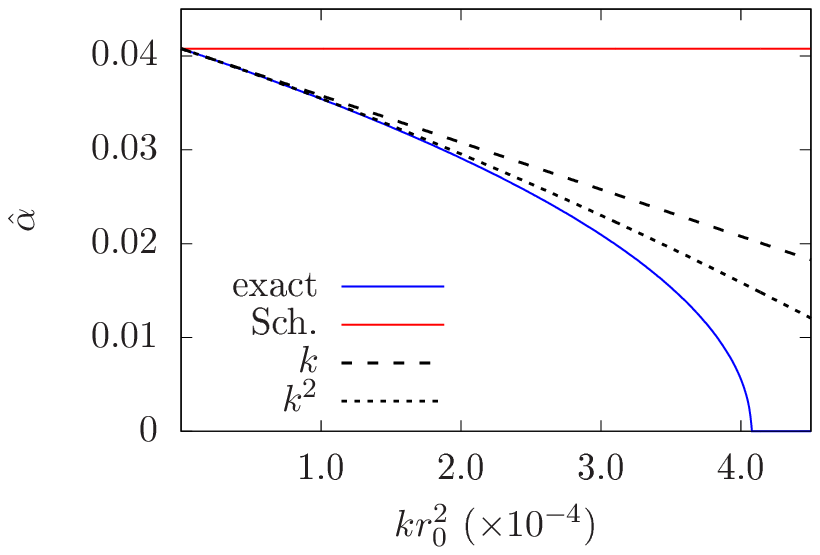}
   \caption{$\hat{\alpha}$ from Eq.~\Eqref{alphak2}.}
   \label{fig_anglesk_compare2}
  \end{subfigure}
  \caption{(Colour online.) Comparison of the approximate bending angle with the exact formula, for the case $m=0.01r_0$, $\gamma=0$, and $\beta=0$. The solid blue line corresponds to the exact bending angle calculated with \Eqref{RIangle}, and the horizontal solid red line is the exact bending angle in the Schwarzschild case. (a) The dashed, dotted, and dashed-dotted curves are calculated from \Eqref{alphak1} plotted up to increasing orders in $m/r_0$. (b) The dashed, dotted, and dashed-dotted curves are calculated from \Eqref{alphak2} plotted up to increasing orders in $k r_0^2$.}
  \label{fig_anglesk_compare}
 \end{center}
\end{figure}

The accuracy of the this approximate result can be compared against the exact expression in \Eqref{RIangle}, as shown in Fig.~\ref{fig_anglesk_compare1}. In Fig.~\ref{fig_anglesk_compare1}, we compare \Eqref{alphak1} to the exact result given in Eq.~\Eqref{RIangle} (shown as the solid curve) for $m=0.01r_0$. The dashed line is the plot of \Eqref{alphak1} keeping up to first order in $m/r_0$ only; hence, we see that the bending angle underestimates the exact result by around $\sim0.0007$. However the rate of change with respect to $k$ follows the exact curve quite closely, as no approximation in $\kappa$ has been made in Eq.~\Eqref{alphak1}. When the higher-order terms are included (dotted line for up to $m^2/r_0^2$ and dash-dotted line for up to $m^3/r_0^3$) the bending angle has excellent agreement with the exact result.\footnote{A similar treatment to the singular perturbation theory  has been applied in various areas of physics, for example, in the strong coupling expansions \cite{Bender:1980effective,Bender:1981strong}, in general mathematical physics \cite{Bender:1994determination}, and in the continuum limit of lattice approximations in boundary-layer theory \cite{Bender:1997continuum,Bender:2002boundary}, just to name a few.}

If the parameter $k$ is small such that $\kappa \ll 1$, the power expansion in $\kappa$ is justifiable,
\begin{align}
 \hat{\alpha}&\sim \frac{4m}{r_0}+\brac{-4 + \frac{15}{4}\pi }\frac{m^2}{r_0^2} + \brac{\frac{98}{3}-\frac{15\pi}{2}}\frac{m^3}{r_0^3}\nonumber\\
  &\quad + \sbrac{-\frac{m}{2r_0} - \brac{\frac{3}{2} - \frac{15\pi}{32}} \frac{m^2}{r_0^2} - \brac{-\frac{5}{4} - \frac{15}{16}\pi + \frac{225}{512}\pi^2} \frac{m^3}{r_0^3}} \kappa \nonumber\\
  &\quad + \sbrac{- \frac{m}{32r_0} + \brac{-\frac{7}{32} + \frac{45}{512}\pi} \frac{m^2}{r_0^2} - \brac{\frac{27}{64} - \frac{135}{256} \pi + \frac{675}{4096}\pi^2} \frac{m^3}{r_0^3}} \kappa^2\nonumber\\
  &\quad +\mathcal{O}\brac{m^4/r_0^4,\kappa^3}.
\label{alphak2}
\end{align}
The comparison between Eqs.~\Eqref{alphak2} and \eqref{RIangle} is shown in Fig.~\ref{fig_anglesk_compare2}. Since the former is also a small-$k$ expansion, we see the expected result that the curve deviates away from the exact result as $k$ increases. As expected, if higher orders of $k$ are included, the deviation is smaller.

The first line on the right-hand side of Eq.~\Eqref{alphak2} is obviously the Schwarszchild light bending in GR. Let us denote it as $\hat\alpha_\mathrm{Sch}$. If we restore $\kappa$ in terms of $k=\frac{\Lambda}{3}$, we obtain that the leading correction due to the cosmological constant is negative,
\begin{align}
 \hat{\alpha}& \sim \hat\alpha_\mathrm{Sch}-\frac{\Lambda r_0^3}{6m},
\end{align}
which is precisely the correction due to the cosmological constant obtained in Ref.~\cite{Rindler:2007zz}. The additional terms in Eq.~\Eqref{alphak2} provide the higher-order corrections of the bending angle.

\subsection{Lensing in the MK spacetime with \texorpdfstring{$k=0$}{k=0}}

We now consider the contribution of the $\gamma$ term in lensing in the MK metric. Thus we now fix $k=0$. As we have seen in Sec.~\ref{effect}, for $\gamma>\gamma_*$, the bending angle is diminished as $\gamma$ increases. Thus we have a similar situation to the Schwarzschild--de Sitter case in which the diminished bending angle causes the path to intersect the optic axis further away from the lens. Beyond $\gamma>\gamma_{\mathrm{crit}}$, the null trajectory passing through $r_0$ no longer intersects the axis before crossing the horizon.

Our approach here is similar to the the Schwarzschild--de Sitter case. To find an approximate expression for the bending angle, we have to perform a small-$m$ and $\gamma$ expansion with care. In this case, we need to expand while still remaining in the $m>m_{\mathrm{crit}}$ region, depicted as the shaded region in Fig.~\ref{fig_gamstar}.

\begin{figure}[t]
 \begin{center}
  \begin{subfigure}[b]{1\textwidth}
  \centering
   \includegraphics[scale=0.8]{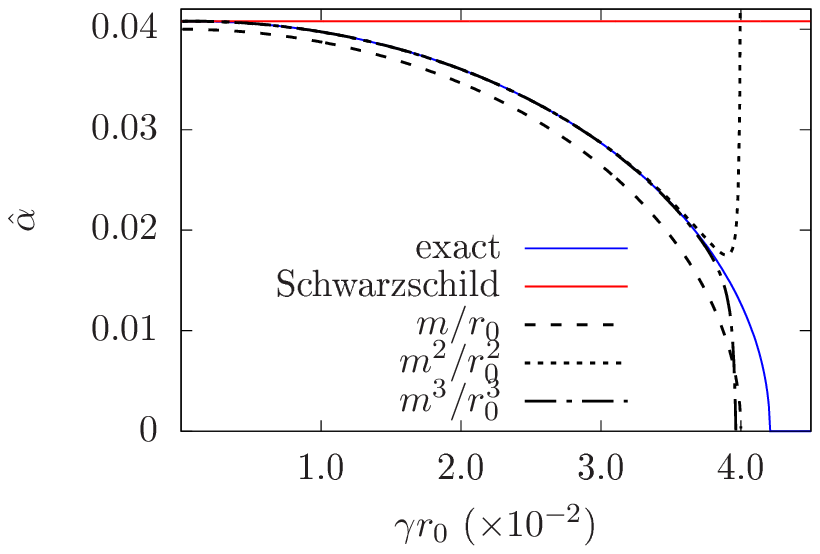}\includegraphics[scale=0.8]{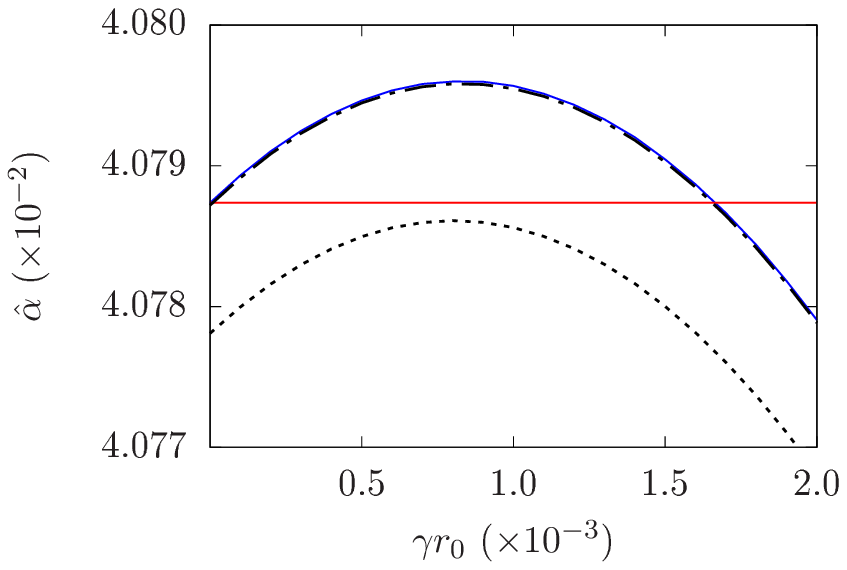}
   \caption{$\hat{\alpha}$ from Eq.~\Eqref{alphagam1}.}
   \label{fig_anglesgam_compare1}
  \end{subfigure}\\
  \begin{subfigure}[b]{1\textwidth}
  \centering
   \includegraphics[scale=0.8]{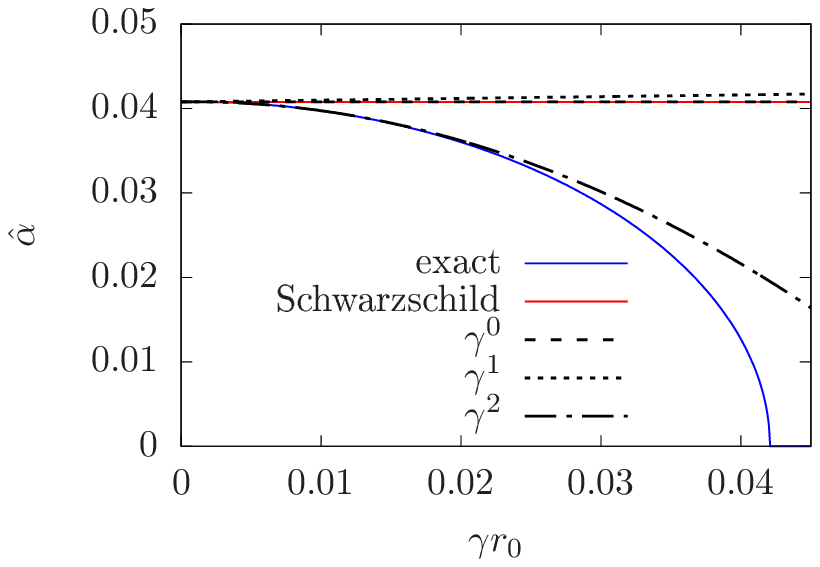}\includegraphics[scale=0.8]{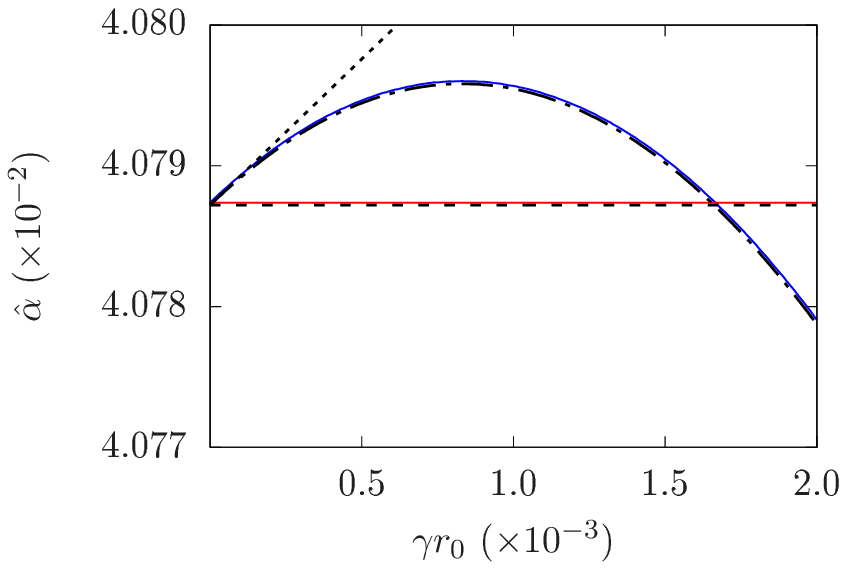}
   \caption{$\hat{\alpha}$ from Eq.~\Eqref{alphagam2}.}
   \label{fig_anglesgam_compare2}
  \end{subfigure}
  \caption{(Colour online.) Comparison of the approximate bending angle with the exact formula, for the case $m=0.01r_0$, $k=0$, and $\beta=0$. The solid blue line corresponds to the exact bending angle calculated with \Eqref{RIangle}, and the horizontal solid red line is the exact bending angle in the Schwarzschild case. (a) The dashed, dotted, and dashed-dotted curves are calculated from \Eqref{alphagam1} plotted up to increasing orders in $m/r_0$. (b) The dashed, dotted, and dashed-dotted curves are calculated from \Eqref{alphagam2} plotted up to increasing orders in $\gamma r_0$. The right panels for each case shows the details at small $\gamma r_0$, which are not visible in the left panels where the full ranges are plotted.}
  \label{fig_anglesgam_compare}
 \end{center}
\end{figure}

Because the leading behaviour of $m_\mathrm{crit} (\gamma)$ in Eq.~\Eqref{eqn:Mcritg} is linear, the parameter $\gamma$ must diminish at the rate of at least as $\gamma \propto m$. Therefore we reparametrise $\gamma$ by setting
\begin{align}
\gamma r_0 \equiv w\frac{m}{r_0},
\label{gamsub}
\end{align}
where $w$ is a dimensionless parameter taken to be of order $\mathcal{O}(1)$ or less. Substituting \Eqref{gamsub} into \Eqref{RIangle} and expanding in powers of $m/r_0$ gives
\begin{align}
 \hat{\alpha}&\sim \sqrt{16-w^2}\frac{m}{r_0} +  \frac{1}{\sqrt{16-w^2}} \brac{-16+15\pi + 8w -w^2 + \frac{1}{2}w^3} \frac{m^2}{r_0^2}\nonumber\\
     &\quad + \frac{1}{\brac{16-w^2}^{3/2}} \left[ \frac{6272}{3} - 480\pi + \brac{-256+240\pi} w \right. \nonumber\\
     &\hspace{1cm} \left. - \brac{176 - 60\pi + \frac{225}{32}\pi^2}w^2 - (-32+30\pi) w^3 - \frac{25}{2} w^4 + w^5 + \frac{1}{3}w^6\right]\frac{m^3}{r_0^3}\nonumber\\
     &\quad +\mathcal{O}\brac{m^4/r_0^4}. \label{alphagam1}
\end{align}
In this case, we see that the expression only holds for $w<4$, which is consistent with the requirement that $w\lesssim\mathcal{O}(1)$. In Fig.~\ref{fig_anglesgam_compare1}, we compare the accuracy of the approximate formula \Eqref{alphagam1} with the exact one given in Eq.~\Eqref{RIangle} for the case $m=0.01r_0$. The solid blue and red lines represent the exact and Schwarzschild bending angles, respectively. The dashed line corresponds to Eq.~\Eqref{alphagam1} plotted only up to leading order, and we see that the curve underestimates the curve by around $\sim m^2/r_0^2$. When the higher-order terms are included (dotted curve for up to $m^2/r_0^2$ and dashed-dotted curve for up to $m^3/r_0^3$), the approximate bending angle has better agreement with the exact results. We can also see that in the approximate bending angles break down at $\gamma r_0=4m/r_0$ due to the factors of $\sqrt{16-w^2}$ in the coefficients, which is consistent with $m_\mathrm{crit} (\gamma)$ in Eq.~\Eqref{eqn:Mcritg}.

If the parameter $\gamma$ is small such that $w\ll1$, we may expand in the powers of $w$,
\begin{align}
 \hat{\alpha}&\sim \frac{4m}{r_0} +\brac{-4+\frac{15}{4}\pi} \frac{m^2}{r_0^2} + \brac{\frac{98}{3} - \frac{15}{2} \pi} \frac{m^3}{r_0^3} + \sbrac{\frac{2m^2}{r_0^2} + \brac{-4 + \frac{15}{4}\pi} \frac{m^3}{r_0^3}} w \nonumber\\
      &\quad + \sbrac{-\frac{m}{8r_0} - \brac{\frac{3}{8} - \frac{15}{128}\pi} \frac{m^2}{r_0^2} - \brac{- \frac{5}{16} - \frac{15}{64}\pi + \frac{225}{2048}\pi^2} \frac{m^3}{r_0^3}} w^2 + \mathcal{O}\brac{m^4/r_0^4,w^3}.\label{alphagam2}
\end{align}
We can also compare this approximate result to the exact one. In Fig.~\ref{fig_anglesgam_compare2}, we plot \Eqref{alphagam2} up to various orders in the case $m=0.01r_0$. Because Eq.~\Eqref{alphagam2} is a perturbative expansion in $\gamma$, we see that it agrees well with the exact angles for small $\gamma$. The right panel of Fig.~\ref{fig_anglesgam_compare2} zooms in on a smaller domain near $\gamma=0$, where we see that the bending angle initially increases with $\gamma$. Here we can see that up to the linear term, the rate of change with respect to $\gamma$ follows the exact value very closely.

\subsection{Lensing in the MK spacetime with \texorpdfstring{$k\neq0$}{k=/=0}}

Knowing the leading behaviour of $m_{\mathrm{crit}}$ in terms of $k$ and $\gamma$,

\begin{align}
\frac{m_\mathrm{crit}(\gamma,k)}{r_0}\sim \frac{1}{4}\sqrt{\gamma^2+4k},
\end{align}
we can now attempt to find the approximate bending angle in the general case where $k\neq0$ and $\gamma\neq0$. Using \Eqref{ksub} and \Eqref{gamsub} in \Eqref{RIangle} and performing an expansion in $m/r_0$, we find
\begin{align}
 \alpha&\sim \sqrt{16-4\kappa-w^2} \frac{m}{r_0}\nonumber\\
       &\quad+ \frac{1}{\sqrt{16-4\kappa-w^2}} \brac{ -16+ 15\pi + 8w - w^2 + \frac{1}{2}w^3 - 4 \kappa + 2\kappa w } \frac{m^2}{r_0^2}\nonumber\\
    &\quad + \frac{1}{\brac{16-4\kappa-w^2}^{3/2}} \biggl[ \frac{6272}{3} -480\pi +\brac{-256 + 240\pi}w + \brac{-176+60\pi - \frac{225}{32}} w^2 \nonumber\\
    &\hspace{2cm} + \brac{32- 30\pi}w^3 -\frac{25}{2} w^4 + w^5 + \frac{1}{3} w^6 \nonumber\\
    &\hspace{2cm} + \brac{-704 + 240\pi -\frac{225}{8}\pi^2 + \brac{192 -120\pi} w  -52 w^2 +4 w^3 +3 w^4 }\kappa   \nonumber\\
    &\hspace{2cm} -8 \brac{1 - w^2}\kappa^2 + \frac{16}{3}\kappa^3 \biggr] \frac{m^3}{r_0^3} + \mathcal{O}\brac{m^4/r_0^4}.
\end{align}

If we further expand in terms of $\kappa$ and $w$, to leading order in each parameter, we have
\begin{align}
 \alpha&\sim\frac{4m}{r_0}+\frac{2m^2w}{r_0^2}-\frac{m\kappa}{2r_0}\nonumber\\
       &=\frac{4m}{r_0}+2m\gamma-\frac{kr_0^3}{2m}.
\end{align}

\section{Conclusion} \label{Conclusion}

In this paper we have derived an exact expression for the bending of light in the SdS and MK spacetime using the Rindler-Ishak method. Special emphasis has been made to obtain the bending angle for a co-aligned source-lens-observer lensing system. By considering numerical and perturbative methods, we found that the $m\to 0$ is a singular limit for a generic non-vanishing $k$ or $\gamma$. This is because for certain ranges of $m$ where $m<m_{\mathrm{crit}}$, there are no null trajectories connecting a co-aligned source and observer. This behaviour guides our approach in finding the correct perturbative expansion for the bending angle, in which we ensure that $m>m_{\mathrm{crit}}$ throughout the analysis.

The exact solutions also reveal a feature that was unnoticed by previous works, namely that for a small range $0<\gamma<\gamma_*$, the deflection angle is actually larger than the Schwarzschild value in General Relativity. Thus, if the value of $\gamma$ obtained by the fitting of galactic rotation curve falls within this range, it would be possible to be consistent with the corresponding observed bending angles.

Since most of the literature pertaining to the Rindler-Ishak angle debate makes use of the small-angle formula, it would be interesting to look for updated results with the improved expressions given in Eq.~\Eqref{RIangle}. This might be especially useful for further applications such as the `vacuole method' considered in \cite{Ishak:2007ea}. With the exact expression there will be no need to keep track of small angles, in addition to the small spacetime parameters. It is worth noting that Ref.~\cite{Simpson:2008jf} has shown that with the appropriate transformation from the SdS to the Friedman-Robertson-Walker in gauge-independent representations, the cosmological constant does not contribute to lensing at the linear regime.

A similar approach can be considered for the CWG, where a natural extension of this work would be to check against observational data using the exact bending angles found above. Previously, this has been done by Cutajar and Adami \cite{Cutajar:2014gfa} using the deflection formulas obtained by \cite{Edery:1997hu,Sultana:2010zz,Cattani:2013dla}. More recently, an analysis inspired by \cite{Ishak:2007ea} was conducted by \cite{Potapov:2016pgr}. Since we have updated these formulas to exact expressions, it would be worthwhile to revisit the observational data using Eq.~\Eqref{InvAngle1}, and further take into account other physical effects considered by \cite{Simpson:2008jf}.

\section*{Acknowledgements}
Q.W.~is grateful to Professor Philip Mannheim for the inspiration and the encouragement. Q.W.~thanks Professor Carl M.~Bender for the discussion on the singular limit in small $m$ perturbation. We thank Dr.~Cindy Ng for her comments and discussion on the field of gravitational lenses in cosmology, and also our anonymous referee for bringing Ref.~\cite{Arakida:2011ty} to our attention.

\appendix
\section{Derivation of Eq.~(\ref{eqn:Mcritk})}
\label{appendix:A}

In this Appendix, we derive the critical mass as a function of $k$ in Eq.~(\ref{eqn:Mcritk}). Recall that here we are considering the case $\gamma=0$, and we set $\beta=0$. A photon trajectory with fixed $r_0$ and $k$ will intersect the optic axis at $r_{\mathrm{obs}}=1/u_{\mathrm{obs}}$, where
\begin{align}
 u_{\mathrm{obs}}=u(\phi_{\mathrm{obs}}). \label{App_uobs}
\end{align}
Generally speaking, $r_{\mathrm{obs}}$ increases as the lens mass $m$ decreases, since we have a weaker gravitational force to pull the photon back towards the optic axis. There will be a critical value $m=m_{\mathrm{crit}}$, where $r_{\mathrm{obs}}=r_{\mathrm{h}}$, where $f(r_{\mathrm{h}})=0$. In other words, $r_{\mathrm{h}}$ is the cosmological horizon corresponding to the larger positive root of $f(r)=0$, given by 
\begin{align}
 r_{\rm h}&= \frac{2}{\sqrt{3k}} \cos\left[ \frac{1}{3} \cos^{-1}\left(-3m\sqrt{3k}\right)\right].
\end{align}
To derive an approximate expression of $m_{\mathrm{crit}}$,  we propose that $m_{\rm crit}(k)$ has a power expansion when the parameter $k r_0^2$ is small, 
\begin{equation}
\frac{m_{\rm crit}(k)}{r_0} = \sum_{n=1}^{\infty} K_n \left(kr_0^2\right)^{n/2},
\label{eqn:mkn}
\end{equation}
where $K_n$ are dimensionless coefficients. There is no constant term in the expansion because $m_{\rm crit}(k)$ diminishes with $k$ as $k\to 0$. This expansion is validated by the consistency of the following asymptotic analysis.

To the leading order,
\begin{equation}
\frac{m_{\rm crit}(k)}{r_0} \sim K_1 \sqrt{k}r_0,
\end{equation}
and the second argument of the incomplete elliptic integral ${\rm F}(p,q)$ in Eq.~(\ref{Deltaphi}) is small,
\begin{equation}
q^2 \equiv  \frac{u_0-u_-}{u_+-u_-} \sim 4 K_1 \sqrt{k}r_0. 
\end{equation} 


When $m=m_{\rm crit}$, the position $\phi_{\mathrm{obs}}$ is located on the cosmological horizon. Therefore, we invert Eq.~\Eqref{App_uobs} to find $\phi_{\mathrm{obs}}=\phi(u_{\mathrm{h}})=\phi_0 +\frac{\pi}{2}$, where $u_{\mathrm{h}}=1/r_{\mathrm{h}}$. For small $kr_0^2$, the first argument of ${\rm F}(p,q)$ has the form
\begin{equation}
\sin p \equiv \left.\sqrt{\frac{(u_+-u_-)(u_0 -u)}{(u_0-u_-)(u_+ -u)}} \right|_{u=u_{\mathrm{h}}}\sim \frac{1}{\sqrt{2}} - \frac{1-3K_1}{2\sqrt{2}} \sqrt{k}r_0.
\label{eqn:pcrit}
\end{equation}
Because $p$ is not small, we must use the asymptotic expansion of the incomplete elliptic integral ${\rm F}(p,q)$ for small $q$ and {\it arbitrary} $p$,
\begin{eqnarray}
{\rm F}(p,q) &\sim&  p + \frac{1}{4}\left(p - \frac{1}{2}\sin2p \right) q^2  + \frac{9}{64}\left(p - \frac{2}{3}\sin2p + \frac{1}{12}\sin4p \right) q^4 \nonumber\\
&&\quad + \frac{25}{256}\left(p - \frac{3}{4}\sin2p + \frac{3}{20}\sin4p - \frac{1}{60}\sin6p \right) q^6 + \mathcal{O}\left( q^8\right).
\label{eqn:Fpq}
\end{eqnarray} 
All together, we have
\begin{equation}
\phi(u_{\rm h}) \sim \phi_0 +\frac{\pi}{2} -(1-2K_1) \sqrt{k}r_0.
\end{equation}
Because at the cosmological horizon $\phi(u_{\mathrm{h}})=\phi_0 +\frac{\pi}{2}$, this leads to
\begin{equation}
K_1=\frac{1}{2}.
\end{equation}

Having determined the leading-order coefficient, we proceed to the sub-leading-order term,
\begin{equation}
\frac{m_{\rm crit}(k)}{r_0} \sim \frac{1}{2} \sqrt{k}r_0 + K_2 kr_0^2.
\end{equation}
Expanding $p$ and $q$ to sub-leading order gives
\begin{eqnarray}
\sin p &\sim& \frac{1}{\sqrt{2}} + \frac{1}{4\sqrt{2}} \sqrt{k}r_0 - \frac{3(7 - 16 K_2)}{32\sqrt{2}} kr_0^2, \nonumber\\
q^2 &\sim& 2 \sqrt{k}r_0 - \left(3 - 4 K_2 \right) kr_0^2. 
\end{eqnarray} 
The critical bending at the cosmological horizon leads to
\begin{equation}
K_2=\frac{1}{2}-\frac{15}{64}\pi.
\end{equation}

Similarly, to the next order,
\begin{equation}
\frac{m_{\rm crit}(k)}{r_0} \sim \frac{1}{2} \sqrt{k}r_0 + \left(\frac{1}{2}-\frac{15}{64}\pi \right) kr_0^2 + K_3 k^{3/2}r_0^3,
\end{equation}
and
\begin{eqnarray}
\sin p &\sim& \frac{1}{\sqrt{2}} + \frac{1}{4\sqrt{2}} \sqrt{k}r_0 + \frac{3(4 - 15\pi )}{128\sqrt{2}} kr_0^2 + \frac{148 + 285\pi + 768 K_3}{512\sqrt{2}} k^{3/2}r_0^3, \nonumber\\
q^2 &\sim& 2 \sqrt{k}r_0 - \left(1 + \frac{15}{16}\pi \right) kr_0^2 + \left(2+ \frac{45}{16}\pi + 4K_3\right)  k^{3/2}r_0^3. 
\end{eqnarray}
By solving the critical bending at the cosmological horizon, we get
\begin{equation}
K_3=\frac{75}{128}\left( -1+\frac{3}{8}\pi\right)\pi.
\end{equation}

Substituting the values of $K_1$, $K_2$, and $K_3$ into Eq.~(\ref{eqn:mkn}), we get Eq.~(\ref{eqn:Mcritk}).

\section{Derivation of Eq.~(\ref{eqn:Mcritg})}
\label{appendix:B}

In this Appendix, we derive the critical mass as a function of $\gamma$ in Eq.~(\ref{eqn:Mcritg}). Recall that we are considering the MK metric with $k=0$ and $\beta=0$. The method is very similar to that in Appendix \ref{appendix:A}, except in this case $m_{\mathrm{crit}}$ corresponds to the limit where $u_{\mathrm{obs}}= 0$. First, we propose that $m_{\rm crit}(\gamma)$ has a power expansion when the parameter $\gamma r_0$ is small,
\begin{equation}
\frac{m_{\rm crit}(\gamma)}{r_0} = \sum_{n=1}^{\infty} \Gamma_n \left(\gamma r_0\right)^n,
\label{eqn:mgn}
\end{equation}
where $\Gamma_n$ are dimensionless coefficients. As in Eq.~(\ref{eqn:mkn}), there is no constant term. This expansion is validated by the consistency of the following asymptotic analysis.

To the leading order,
\begin{equation}
\frac{m_{\rm crit}(\gamma)}{r_0} \sim \Gamma_1 \gamma r_0,
\end{equation}
and the second argument of the incomplete elliptic integral ${\rm F}(p,q)$ in Eq.~(\ref{Deltaphi}) is small,
\begin{equation}
q^2 \equiv  \frac{u_0-u_-}{u_+-u_-} \sim 4 \Gamma_1 \gamma r_0.
\end{equation}

Because $k=0$ in this calculation, there is no cosmological horizon. When $m=m_{\rm crit}$, we have $\phi(u_{\mathrm{obs}}=0)=\phi_0 +\frac{\pi}{2}$. For small $\gamma r_0$, the first argument of ${\rm F}(p,q)$ has the form
\begin{equation}
\sin p \equiv \left.\sqrt{\frac{(u_+-u_-)(u_0 -u)}{(u_0-u_-)(u_+ -u)}} \right|_{u=0}\sim \frac{1}{\sqrt{2}} - \frac{1-6 \Gamma_1}{4\sqrt{2}} \gamma r_0.
\end{equation}
Again the same leading term as in Eq.~(\ref{eqn:pcrit}) indicates the critical bending. Using the same asymptotic expansion of ${\rm F}(p,q)$ in Eq.~(\ref{eqn:Fpq}), we get
\begin{equation}
\phi(u=0) \sim \phi_0 +\frac{\pi}{2} - \left( \frac{1}{2}- 2 \Gamma_1\right) \gamma r_0.
\end{equation}
The critical bending leads to
\begin{equation}
\Gamma_1=\frac{1}{4}.
\end{equation}

To sub-leading order,
\begin{equation}
\frac{m_{\rm crit}(\gamma)}{r_0} \sim \frac{1}{4} \gamma r_0 + \Gamma_2 \gamma^2 r_0^2,
\end{equation}
and
\begin{eqnarray}
\sin p &\sim& \frac{1}{\sqrt{2}} + \frac{1}{8\sqrt{2}} \gamma r_0 + \frac{(19 +192 \Gamma_2)}{128\sqrt{2}} \gamma^2 r_0^2, \nonumber\\
q^2 &\sim& \gamma r_0 - \left( \frac{1}{4} - 4 \Gamma_2 \right) \gamma^2 r_0^2.
\end{eqnarray}
The critical bending leads to
\begin{equation}
\Gamma_2=-\frac{1}{8}-\frac{15}{256}\pi.
\end{equation}

Similarly, to the next order,
\begin{equation}
\frac{m_{\rm crit}(\gamma)}{r_0} \sim \frac{1}{4} \gamma r_0 - \left(\frac{1}{8} + \frac{15}{256}\pi \right) \gamma^2 r_0^2 + \Gamma_3 \gamma^3 r_0^3,
\end{equation}
and
\begin{eqnarray}
\sin p &\sim& \frac{1}{\sqrt{2}} + \frac{1}{8\sqrt{2}} \gamma r_0 - \frac{5(4 + 9\pi)}{512\sqrt{2}} \gamma^2 r_0^2 + \frac{-76 + 105\pi + 6144 \Gamma_3}{4096\sqrt{2}} \gamma^3 r_0^3, \nonumber\\
q^2 &\sim& \gamma r_0 - 3\left(\frac{1}{4} + \frac{5}{64}\pi \right)\gamma^2 r_0^2 + \left(\frac{3}{4}+ \frac{15}{64}\pi + 4\Gamma_3\right) \gamma^3 r_0^3.
\end{eqnarray}
By solving the critical bending at $u_{\mathrm{obs}}=0$, we get
\begin{equation}
\Gamma_3=\frac{1}{16} +\frac{15}{1024}\pi +\frac{225}{8192}\pi^2.
\end{equation}

Plugging in the values of $\Gamma_1$, $\Gamma_2$, and $\Gamma_3$ to the expansion in Eq.~(\ref{eqn:mgn}), we get Eq.~(\ref{eqn:Mcritg}).

\bibliographystyle{cglens}

\bibliography{cglens}

\end{document}